# Structure of surfaces and interfaces of Poly(*N*,*N*-dimethylacrylamide) hydrogels

Guillaume Sudre,[a,‡] Dominique Hourdet,[a] Fabrice Cousin,[b] Costantino Creton[a]

and Yvette Tran[a]

[a] *Laboratoire de Sciences et Ingénierie de la Matière Molle, UMR 7615 CNRS/UPMC/ESPCI ParisTech, 10 rue Vauquelin, F-75231 Paris Cedex 5, France.*

[b] *Laboratoire Léon Brillouin, CEA-CNRS, Saclay 91191 Gif-sur-Yvette Cedex, France*

[‡] *Current address: Materials Sciences Division, Lawrence Berkeley National Laboratory, University of California, Berkeley, CA 94720, USA*

*Corresponding author e-mail: yvette.tran@espci.fr

# ABSTRACT

We investigated the surface structure of hydrogels of poly(*N,N*-dimethylacrylamide) (PDMA) hydrogels synthesized and crosslinked simultaneously by redox free radical polymerization. We demonstrate the existence of a less crosslinked layer at the surface of the gel at least at two different length scales characterized by shear rheology and by neutron reflectivity, suggesting the existence of a gradient in crosslinking. The thickness and composition of the layer is shown to depend on the degree of hydrophobicity of the mold surface and is thicker for more hydrophobic molds. While the macroscopic tests proved the existence of a relatively thick undercrosslinked layer, we also demonstrated by neutron reflectivity that the gel surface at the submicrometric scale (500 nm) was also affected by the surface composition of the mold. These results should have important implications for the measurement of macroscopic surface properties of these hydrogels such as friction or adhesion.

INTRODUCTION

Hydrogels are ubiquitous in biological systems and biomedical applications, as macro-objects such as contact lenses for instance, or as micro- or nano-objects such as carriers for drug delivery.[1-5] Often the required property is that of a molecular sponge able to store specific molecules and release them on cue and the low elastic modulus of gels, comparable to that of living tissues is often appealing for these applications. Yet in many cases gels are not suspended in a solution but need to be in contact with other gels or with hard surfaces. The question of how to control adhesion and friction of hydrogels on surfaces becomes then important.

An essential part of the puzzle is the understanding of the interfacial or surface structure of the gels and its relation with macroscopic properties. For polymer melts and polymer glasses, this combined study of polymer interfaces and macroscopic adhesion has proved an extremely valuable tool to understand the adhesive mechanisms from a more molecular standpoint.[6, 7] Although the methodology used in these cases can be transposed to hydrogels, the details of the chemistry and polymer physics are quite different.

The main component of hydrogels is water which by itself cannot sustain stress. Therefore any stress to be transferred across a gel/surface interface has to be through the polymer chains. Yet these polymer chains are diluted in water (typically around 10% polymer) reducing the strength of the interactions which are in addition screened by the very high dielectric constant of water. This dilution suggests that entanglements, that play a major role in interfaces between polymer melts and polymer glasses, may not be as important for gel interfaces while the effect of specific interactions (H-bonding, acid-base, electrostatic...) may become more important. This stresses the need to understand how the polymer chains are organized near a surface when immersed in water. On the theoretical side, Joanny et al.[8] have studied the behavior of gels from a fundamental point of view, but they focused mostly on the competition between surface tension effects and elasticity and did not consider the details of the gel chemistry. A more chemically oriented study was carried out by the Peppas group[9, 10] who studied the effect of the addition of tethered polymer chains at the interface on the adhesion between two hydrogels and showed that PEG-grafted poly(methacrylic acid) hydrogels were interesting mucoadhesive carriers since the PEG chains can interpenetrate the gel mucus layer of the small intestine.

A very interesting experimental investigation combining the structure of gel interfaces and the macroscopic frictional properties was carried out by Gong and her group.[11, 12] They discovered that when the chemical hydrogel was synthesized by UV free radical polymerization, the chemical composition of the surface of the mold had a profound influence on the frictional properties of the gel surface after demolding.[13-15] With a series of elegant experiments they demonstrated that the presence of oxygen adsorbed on the surface of the mold when this surface was hydrophobic, caused a slightly lower polymerization rate than in the bulk. When the polymerization reached a gel point, this difference was amplified and resulted in the end in the formation of an undercrosslinked hydrogel layer near its surface over a distance which could be up to a mm[13, 16-18]. This peculiar result clearly affected the friction properties of the gel but may be due to the very slow UV polymerization process that they used.

However many hydrogels are synthesized by simple redox polymerization and in order to investigate their adhesion properties, it is essential to carefully study the structure of the surfaces and interfaces at a macroscopic and at a local scale since one can suspect that polymer-polymer entanglements, specific interactions and generally speaking, interpenetration between gels can only occur over a distance of the order of the mesh size, i.e. a few nm. We therefore carried out two separate sets of interface characterization experiments targeted at two different length scales. At the macroscopic scale, we compared for the same gels, the elastic modulus in compression of cylinders measured at large strains, which should be representative of bulk properties, and the shear modulus of disks measured in a rheometer at small strains, which should be very sensitive to the presence of surface layers.

At the molecular scale, the structure of the interface between a model hydrogel and a polymer brush grafted covalently on a silicon surface was investigated by neutron reflectivity. The choice of the specific gel and brush was done in order to be able to use the exact same model system to carry out adhesion tests as a function of pH and temperature. The adhesion results and methodology will be presented in a separate paper while this paper will focus on the structure of the interface.

## EXPERIMENTAL SECTION

**Preparation of hydrogels**

The synthesis of poly(*N,N*-dimethylacrylamide) hydrogels has been carried out by free radical polymerization using *N,N'*-methylene-bis-acrylamide (MBA, Sigma-Aldrich, 99%) as cross-linker. After dissolution of *N,N*-dimethylacrylamide (DMA, Aldrich, 99%), MBA and potassium persulfate (KPS, Sigma-Aldrich, >99%) in Milli-Q water, the solution was deoxygenated with a bubbling of nitrogen during 30 min. After a rapid addition of N,N,N',N'-tetramethylethylenediamine (TEMED, Sigma-Aldrich, 99%) under stirring, the solution was then transferred to a hydrophobically treated mold placed under a nitrogen atmosphere. The red-ox initiation rapidly took place[19] and the polymerization was let to proceed during 4 h. Then, the mold was opened and the gel was immersed in Milli-Q water for dialysis. Water is changed twice a day during one week and the hydrogel was finally stored in its swollen state until final use. The analysis of extractible was not systematic, but no residual monomer was detected by size-exclusion chromatography (SEC) after extraction and the sol fraction was not measurable or below 2 wt% of the total network. This result is consistent with the work of Gundogan *et al.*[20]

We chose to work with normalized gels. The mass ratio of DMA in water was 10 wt% while the molar ratio of cross-linker (MBA) to monomer was varied from 0.5 to 2 mol%. The quantity of initiator (KPS and TEMED) was set to 1 mol% of the monomer quantity.

For compression tests, cylindrical gels were formed in home-made polydimethylsiloxane (PDMS) molds (diameter: 6-8 mm, height: 10 mm). For rheology and reflectivity experiments, plate-shaped gel samples were specifically prepared by molding between two glass slides (100×75×2.5 mm$^3$) as drawn in Figure 1. In order to improve the removal of the gel from the mold, glass slides were treated hydrophobic by silanization, as described in the following part. After being modified, the various substrates were kept in the dark at room temperature.

For rheology, discs with a diameter of 28 mm were punched out of the gel plates. For neutron reflectivity, the dialysis step was carried out in deuterium oxide equilibrated at pH 9 and then rectangles (80×40 mm$^2$) were cut out of the gel plates to fit both the dimensions of the silicon monocrystal and of the liquid trough used in these experiments. Some plate-shaped samples were also synthesized in polypropylene (PP) Petri dishes, open to a nitrogen atmosphere. With all these hydrophobic treatments, the demolding of the gel happened without causing any damage to the surfaces of the gels.

**Modification of the surfaces of glass and silicon substrates**

In order to improve the removal of the gel from the mold, glass slides were treated hydrophobic by silanization. Three different hydrophobic silanes were used for that purpose (see Figure 1): hexamethyldisilazane (HMDZ, Gelest, 98.5%) is a small molecule which passivates the surface with methyl groups; octadecyltrichlorosilane (OTS, Gelest, 97%) corresponds to aliphatic saturated chains of 18 carbons and gives access to more hydrophobic surfaces than HMDZ; finally 1H,1H,2H,2H-perfluorodecyltrichlorosilane (FTS, Gelest, 97%) is shorter than OTS but its chain is fluorinated and its grafting gives rise to the most hydrophobic surfaces.

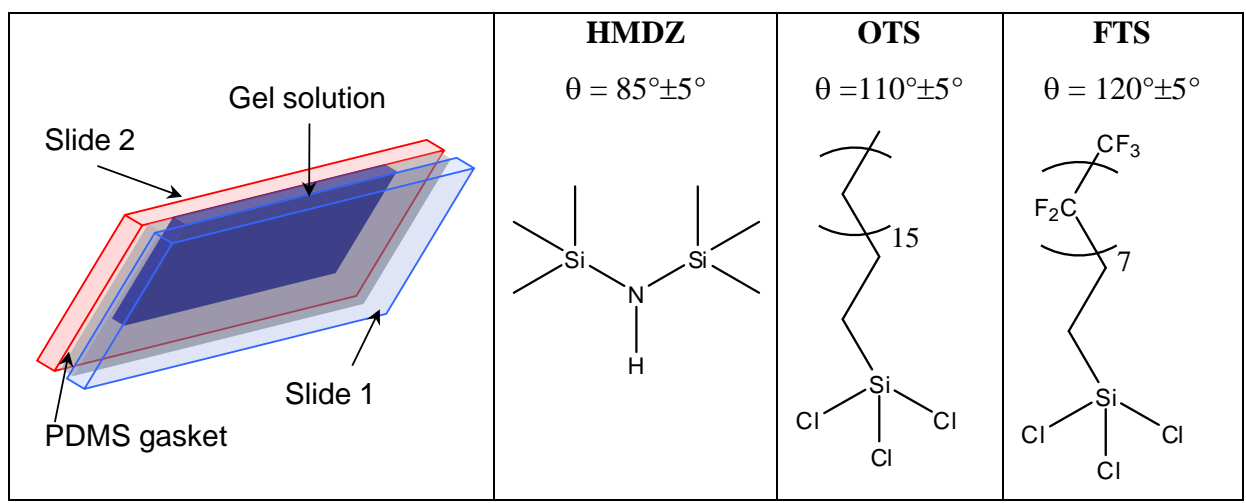

*Figure 1. Schematic representation of the typical mold used for the synthesis of gel plates and silane molecules used for the hydrophobic treatment of glass plates; with θ their contact angle.*

Prior to any surface modification, the glass substrate was cleaned and rejuvenated by an immersion in an active "piranha" solution (70 vol% of sulfuric acid (96%), 30 vol% of hydrogen peroxide (30 wt% in water), heated at 150 °C), extensively rinsed with Milli-Q water and dried with a nitrogen flow. For HMDZ, the surface modification was carried out with a vapor phase method: after rejuvenation, glass substrates were placed in a close environment with a few droplets of HMDZ for 2 hours; the glass plates were then removed from the reactor and rinsed with ethanol and water. The high contact angle on the glass observed during rinsing and the absence of remaining pinned droplets on the surface without any drying are the evidence of a successful modification. For OTS[21] and FTS, the silanization was performed in toluene with a concentration of 0.15 vol% during 2 hours at room temperature. The glass plates were then extensively rinsed with toluene and sonicated prior to drying with a gentle flow of nitrogen. The contact angle with water was finally evaluated to ensure the quality of the hydrophobic modification. The contact angle varied from 85° on HMDZ-treated to 120° on FTS-treated glass plates.[22]

To avoid specific interactions between the gels and the silanol groups located at the surface of the silicon wafers,[23] the wafers were treated with a poly(acrylic acid) (PAA) brush covalently grafted on the surface. At pH > 5 the acrylic acid is charged and one does not expect any interactions between the gel and the surface.[24] The synthesis of the PAA brush has recently been published elsewhere[25] and the synthesis method that we used is briefly described in the supporting information.

**Compression tests**

The compression tests have been carried out on a home-made force apparatus.[26] During the experiment, force and displacement were measured as a function of time, with a resolution of 0.001 N and 0.1 μm respectively. From this measurement, the small strain Young's modulus can be obtained. The undeformed samples have a typical height $h_0$ of at least 5 mm and a diameter of 7 to 8 mm from which can be calculated the cross-section $S_0$. The flat faces of the cylinders were cut if needed to have the most flat and regular possible faces perpendicular to the axis of compression. The height and diameter of the sample were then measured within a precision of 0.1 mm. The compression test was realized with the sample placed between two glass slides. The gel (at swelling equilibrium) was covered with dodecane to limit the evaporation of water and to ensure a perfect lubrication at the glass/gel interface and to prevent the cylinder from barrelling, which was essential to carry out an ideal uniaxial compression up to large strains.[27, 28] At the beginning of each experiment, the cylinder faces were put in contact with the two glass slides, applying a pre-load of 30 mN. Then, the compression test was carried out. The other details of the measurement and of the analysis of the data are as previously reported.[26]

**Rheology**

To investigate the dissipative properties of the hydrogels, their linear viscoelastic properties have been studied on a strain-controlled rheometer Rheometrics RFSII. We used a parallel-plate geometry with a 25 mm diameter. Both plates were roughened to avoid any slippage of the gels at the contact with the rheometer tools. The gel sample was carefully deposited on the bottom plate of the rheometer and the top plate of the rheometer was moved down and put in contact with the gel to read a normal force of 0.1 N. Then, the samples were immersed in water to avoid drying.

Two types of measurements have been carried out: the first one is a strain sweep from 0.045% to 10% with a frequency equal to 1 Hz; the second one is a frequency sweep between 0.1 Hz and 20 Hz, with a fixed strain (0.1%).

The strain sweep assures that the frequency sweep is performed in the linear region and that there is no slippage at the interface between the gel and the plates of the rheometer. A second strain sweep was realized over a shorter range to make sure that the first measure did not alter the gel sample. For a strain lower than a few percents, the storage modulus remains constant, and decreases only for a higher strain value; this is usually the signature of damage occurring in these conditions. With the second measurement, we made sure the system has not been damaged during the first measurement since no significant variation of the moduli could be observed.

The frequency sweeps were then performed in the linear strain regime of the material.

**Neutron reflectivity**

Neutron reflectivity measurements were performed at silicon-liquid interface on the reflectometer EROS at the Laboratoire Léon Brillouin, CEA-Saclay (France). We used protonated gels swollen in deuterated water in order to enhance the contrast between the polymer and the immersing medium. In this paper, we focus on the position of the total reflection plateau. Neutron reflectivity is sensitive to the profile of the neutron scattering density between two semi-infinite media separated by a "flat" interface. When arriving from the semi-infinite medium 1 (silicon here), the neutron beam is reflected off the semi-infinite medium 2 (PDMA gel in our case) which has a higher neutron scattering density $\rho_2 > \rho_1$. The position of the total reflection plateau is specific to the difference in neutron scattering density between the two semi-infinite media. The total reflection plateau is obtained for wave vectors below the critical wave vector $q_c$, which can be written as:

$$q_c = \frac{2\pi}{\lambda_c}\sin\theta_c = \sqrt{4\pi(\rho_2 - \rho_1)}.$$

As for silicon, the neutron scattering density is well known and equal to $\rho_1 = 2.08\,10^{-6}$ Å$^{-2}$. Determining the critical wave vector for silicon wafers in contact with identical gels being synthesized in various molds gives directly access to the polymer concentration in the gel near the surface of the silicon substrate. PDMA hydrogels have been synthesized in deuterium oxide instead of water as described previously and dialyzed in the same deuterium oxide buffer at pH 9. The silicon wafers were modified with PAA bimodal brushes: 23.6 kg.mol$^{-1}$ PAA chains were grafted with a density of 1.35 10$^{-1}$ nm$^{-2}$ and 2.36 kg.mol$^{-1}$ PAA chains were grafted with a density of 6 10$^{-2}$ nm$^{-2}$ ± 1.5 10$^{-2}$ nm$^{-2}$, as presented in the supporting information.

It should be noted that the two media are considered as semi-infinite if their thickness is higher than the penetration depth of the evanescent wave generated by the total reflection. The optical penetration depth is given by:

$$\delta = \frac{\lambda}{4\pi\sqrt{n_1^2 \sin^2\theta - n_2^2}} = \frac{\lambda}{4\pi n_2 \sqrt{\frac{\sin^2\theta}{\sin^2\theta_C} - 1}}$$

where $n_2$ is the neutron refractive index of the medium 2. The condition for total reflection is $\theta \geq \theta_C$ with $\sin\theta_C = \frac{n_2}{n_1}$. The neutron refractive index $n$ is related to the neutron scattering length density $\rho$: $n = 1 - \frac{\lambda^2}{2\pi}\rho.$

In the absence of absorption, the penetration depth becomes infinite at the critical angle of total reflexion. Here, the penetration depth is estimated to 500 nm.[29]

For each gel/brush pair, two types of reflectivity experiments, represented on Figure 2, were carried out: (A) in the first one, the gel, immersed in deuterium oxide, was put on top of the silicon wafer and confined in an adapted chamber to avoid evaporation/exchange; (B) in the second one, a load was added on top of the gel so that it was compressed of about 10 to 20 %.

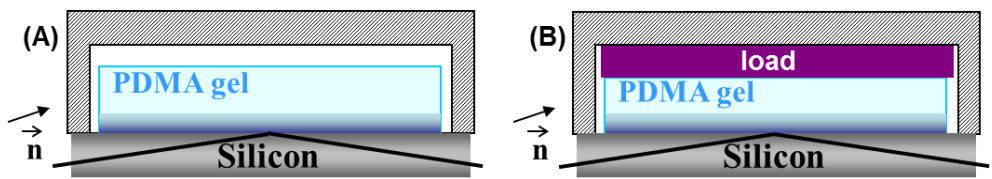

*Figure 2. Set up for neutron reflectivity experiments. (A): the gel sample is put on top of the silicon wafer in the presence of deuterium oxide in a closed chamber; (B): the gel sample is compressed of about 10% on the silicon wafer in the presence of deuterium oxide in a closed chamber.*

RESULTS

**Mechanics: lubricated compression test**

All gels deformed elastically (hysteresis free) as previously reported[30] and showed a linear behavior in compression for the true stress vs. $\lambda$, following the prediction of rubber elasticity behavior in the range up to 30% compression as shown on Figure 3.

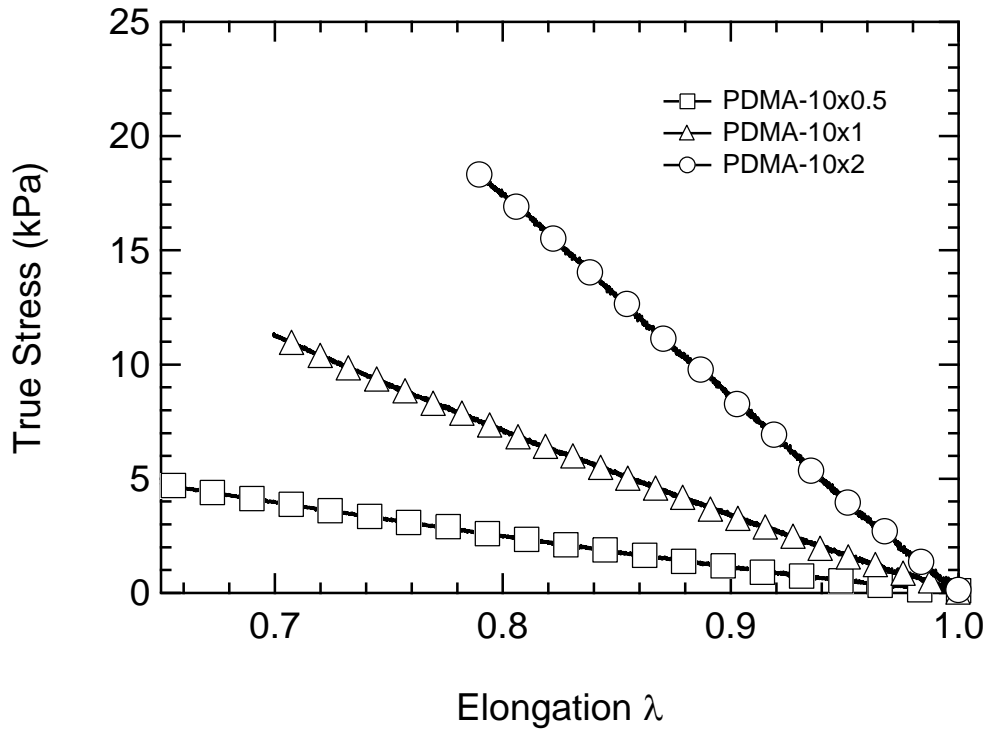

*Figure 3. True Stress vs. elongation data for lubricated compression tests of three different gels with 10 wt% monomer. 0.5 mol% crosslinker ($\square$), 1 mol% crosslinker ($\triangle$) and 2 mol% crosslinker ($\bigcirc$).*

For the surface and interface study we used the gel with 2 mol% crosslinker and the shear modulus *G* obtained from these compression tests by assuming incompressibility and frictionless slippage at the interface, is $29 \pm 3$ kPa.

**Rheology**

The same PDMA hydrogel (10 wt% monomer, 2 mol% crosslinker, swollen to equilibrium) was prepared between different silanized glass substrates used for molding during the gel polymerization (OTS, FTS and HMDZ). A fourth synthesis was also carried out by pouring the reaction medium into a polypropylene (PP) Petri dish while the top surface was in contact with nitrogen atmosphere. As shown in Figure 4, there is a real impact of the surface chemistry of the mold on the measured dynamic storage modulus G', which is frequency independent in the measured range. By comparison with the static reference value determined by compression (G = 29 kPa), the storage modulus obtained by rheology is much smaller and appears to scale with the hydrophilic character of the mold surface: the storage modulus increases when the mold becomes more hydrophilic from 5.2 kPa for the FTS-treated surface to 8.3 kPa for the HMDZ-treated mold.

For the Air-PP mold, this analysis is a bit twisted since the two surfaces of the mold are different: PP has a contact angle equal to $\theta = 100° \pm 5°$ but air or more exactly nitrogen is probably the most hydrophobic medium since its surface tension with water is very high: $\gamma_{air-water} = 72.8 \, mJ.m^{-2}$.

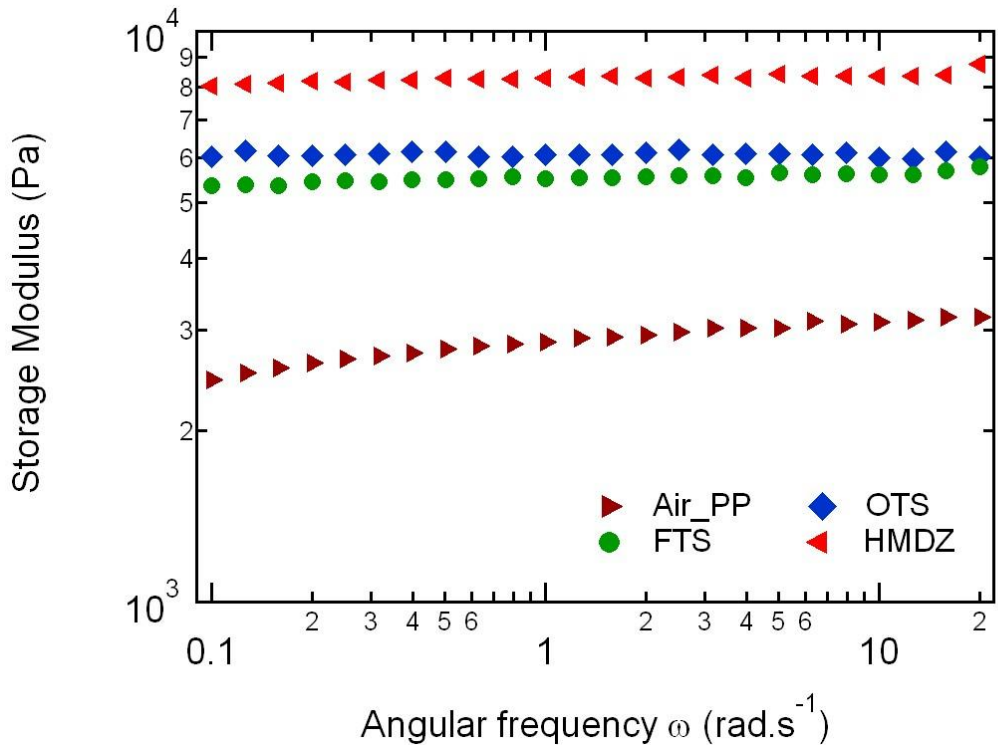

*Figure 4. Storage modulus as a function of angular frequency for the same PDMA-10x2 hydrogel molded between substrates functionalized with different molecules: ♦ OTS; ► Air and PP; ◄ HMDZ; ● FTS.*

**Characterization of the molecular structure at the interface**

To characterize the extent of the concentration depletion at the surface at the molecular scale, we have performed neutron reflectivity experiments with the same PDMA gels synthesized in deuterium oxide. The neutron reflectivity curves obtained are given in Figure 5. We have added the best fit to the experimental data, and we have also plotted a simulation, assuming that the gel was ideally and uniformly swollen in the bulk and at the surface[1]. It should be noted that the PDMA gels were put in contact with silicon wafers which were treated with a poly(acrylic acid) brush. This surface treatment allowed the prevention from specific interactions between the PDMA gels and the silanol groups located at the bare surface of silicon wafers.[23] Here, we focused on the position of the critical wave vector which is specific of the scattering length density of the swollen hydrogel in contact with the silicon wafer functionalized by the poly(acrylic acid) brush. The whole reflectivity curve and the corresponding volume fraction profile that best fit the reflectivity curve are given in the Supporting Information. Actually, the density profile corresponds essentially to the one of the polymer brush since the reflectivity signal is dominated by the contribution of the brush on the first 500 nm near the interface (for q range above $q_C$) but definitely has no effect on the position of $q_C$ as it is about 10 nm-thick.

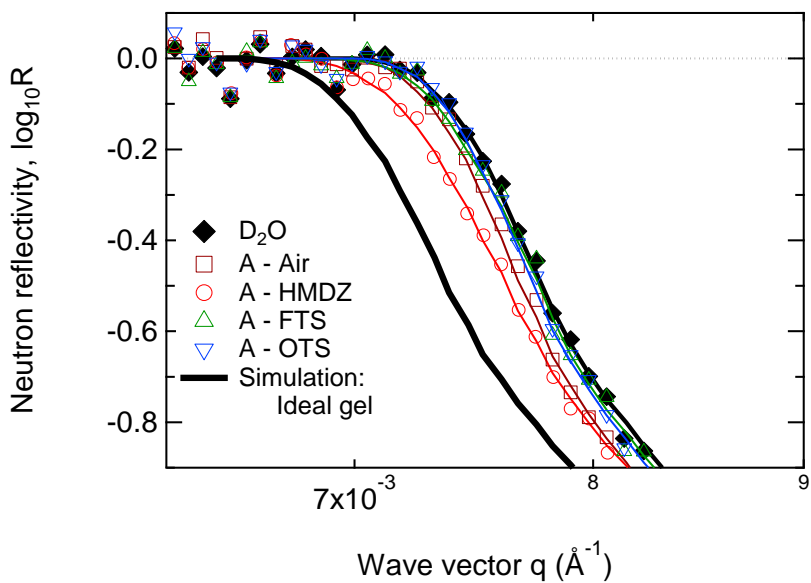

---

[1] The brush conformation is supposed to be identical to experiment B – OTS in this simulation where only the index of the second semi-infinite medium has been modified. The same procedure with any other experiment would have lead to a similar result.

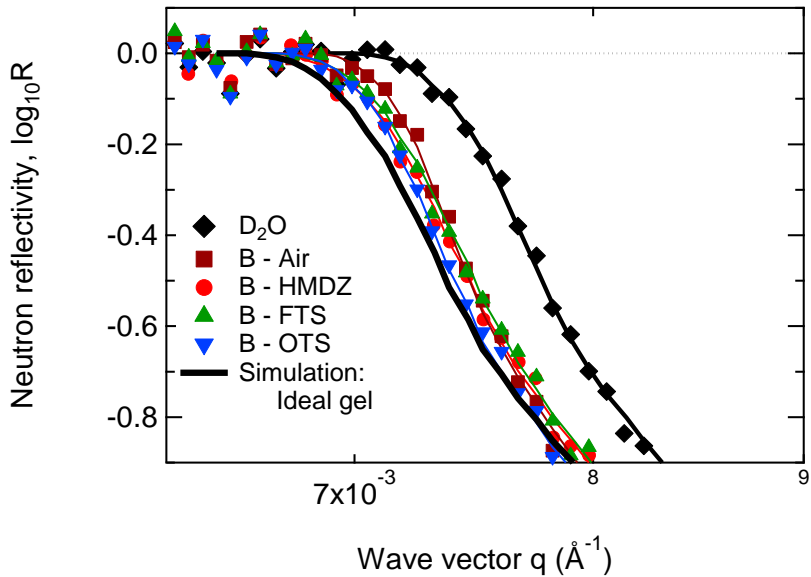

*Figure 5*. *Neutron reflectivity curves, near the total reflection plateau. Open symbols correspond to experiments A and filled symbols to experiments B. The experimental data (dots) are shown with their best fit (corresponding line). The second semi-infinite medium is either water (black), or equilibrated PDMA gels synthesized with a free surface or with the mold surfaces silanized with HMDZ, FTS or OTS. The bold black line corresponds to an ideal PDMA gel as second semi-infinite medium.*

In Figure 5-top, are shown the data from experiments A. The curves obtained from the gel synthesized in contact with air and OTS- and FTS- modified surfaces are very close to the reflectivity curve obtained in the presence of pure deuterium oxide. This means that for gels synthesized in those conditions (against surfaces such as FTS- or OTS-modified glass substrate, or with a free surface) and simply put in contact with the silicon surface, a polymer depleted layer forms near their surface. The average volume fractions of polymer have been estimated from the simulation and the results are summarized in Table 1. For gel synthesized against FTS- or OTS-modified glass substrate, or with a free surface, the volume fractions of gel at the surface range from 0.4% to 1.5% whereas the polymer volume fraction in the bulk of the gel is equal to 7.1% which corresponds to the swelling at equilibrium $Q_e$ of the same PDMA hydrogel determined as $Q_e = 14$. However the gel synthesized in an HMDZ-modified mold behaves differently and the data show that the surface polymer density is much higher at 3.8%.

Data from experiments B are shown on Figure 5-bottom. In this case, the hydrogels are compressed on the silicon substrate. For all surfaces, the total reflection plateau is close to the limit that should be observed for a uniformly swollen gel. The mean polymer density in the first 500 nm exceeds 5.1% and becomes close to – but remains below – the polymer density in the ideal gel, which is equal to 7.1%. By comparing the results from experiments B with those from experiments A, the local polymer concentration near the interface increases regardless of the preparation conditions when the gel is compressed in contact and it gets close to the average bulk composition[2].

|  | Experiment A | | Experiment B | |
| --- | --- | --- | --- | --- |
| Synthesis condition | $(\rho_2 - \rho_1)10^6$ (Å$^{-2}$) | $\phi_{pol}$ and $Q$ | $(\rho_2 - \rho_1)10^6$ (Å$^{-2}$) | $\phi_{pol}$ and $Q$ |
| D$_2$O, pH 2 | 4.38 | $\phi_{pol} = 0\%$ | | |
| Air | 4.30 | $\phi_{pol} = 1.5\%$; $Q = 69$ | 4.10 | $\phi_{pol} = 5.1\%$; $Q = 19$ |
| HMDZ | 4.17 | $\phi_{pol} = 3.8\%$; $Q = 26$ | 4.02 | $\phi_{pol} = 6.5\%$; $Q = 15$ |
| FTS | 4.34 | $\phi_{pol} = 0.7\%$; $Q = 138$ | 4.05 | $\phi_{pol} = 6.0\%$; $Q = 17$ |
| OTS | 4.36 | $\phi_{pol} = 0.4\%$; $Q = 275$ | 4.00 | $\phi_{pol} = 6.9\%$; $Q = 14$ |
| Ideal gel | 3.99 | $\phi_{pol} = 7.1\%$; $Q_e = 14$ | | |

***Table 1.*** *Polymer concentration at interfaces $\phi_{pol}$ and corresponding local swelling $Q$ as a function of the mold surface for PDMA-10x2 hydrogels.*

---

[2] However, since the compression method did not allow a perfect reproducibility, the small variations observed within the experiments B cannot be discussed.

DISCUSSION

From rheology and neutron reflectivity experiments, we have shown that the mold used for the synthesis of the gel has a pronounced effect on its surface morphology. These effects appear to occur at two separate length scales: a macroscopic length scale detectable by rheology and a molecular one detectable by neutron reflectivity. Such an effect of the mold surface on the macroscopic surface properties of the gel (namely friction) has been shown before for a UV polymerization where spatial heterogeneities in reaction rates are more common, but to the best of our knowledge it is the first time that a difference in polymer composition near the surface is directly measured by neutron reflectivity, and in the case of a redox polymerization of a gel.

Gong and her group[11, 12] (and references therein) showed that the surface swelling of their gels was higher if the mold surface was hydrophobic than if it was hydrophilic. Such a difference in swelling and in friction behavior suggested the existence of a surface layer with a lower concentration in polymer and/or crosslinker. The existence of such a layer in their samples was demonstrated by elegant optical experiments.[13, 18, 31]

They first attributed this difference to the high interfacial energy between the gel solution before synthesis and the hydrophobic substrate,[13] leading to oxygen pockets remaining pinned on the substrate to reduce the interfacial energy. The presence of this oxygen would locally modify the polymerization process gradually leading to a gradient in monomer and eventually to the formation of an undercrosslinked and more diluted gel layer near the surface over distances that could reach one mm.[15]

In our system we have clearly reproduced the same effect with a redox polymerization which should not in principle lead to heterogeneities through the thickness of the sample and with a reaction time of minutes rather than hours. Furthermore our neutron reflectivity measurements show that the effect is not only macroscopic but also present at the molecular length scale, and that molds with hydrophobic surfaces lead to a lower average concentration of polymer in the submicrometric vicinity of the surface, evaluated to be the first 500 nm of the hydrogel films. Interestingly even the least hydrophobic mold surface (HMDZ) leads to a polymer depletion near the surface by neutron reflectivity and to a difference in shear modulus between the bulk value (measured by large strain compression) and the more surface-sensitive value measured by shear rheology.

These results suggest that unless special precautions are taken, all gels synthesized by simultaneous polymerization and crosslinking from the monomer, will have a surface driven compositional gradient. Furthermore the neutron reflectivity data show that the extreme surface composition maybe also very sensitive to the details of the synthesis process.

If this scenario is correct, working in a glove box with modified glass substrates stored under a nitrogen atmosphere for a few days could remove the effect of the substrate.

# CONCLUSION

The influence of the mold used during the synthesis by redox free radical polymerization of a chemical gel on the structure of its free surface has been investigated. Mechanical measurements were used to demonstrate the influence of the mold surface on the gel rheological response. A large discrepancy was found between the dynamic storage modulus obtained in rheology on discs (small strains in shear) and the modulus obtained with uniaxial compression tests on cylinders at larger strains. This suggested strongly that the composition of the gel (monomer and crosslinker) near the surface over macroscopic distances was different from that of the bulk.

Furthermore we showed directly by neutron reflectivity that the first 500 nm at the surface of the gel had a different polymer volume fraction than the bulk and were also influenced by the surface composition of the mold.

Our results have shown that among various hydrophobic treatments used to modify the glass molds to allow easy demolding of the gel, the most hydrophilic treatment, HMDZ, gave the results with the lowest discrepancy between the bulk and the surface composition of the gel.

Finally while the characteristics of the free surface of the gel are essential for adhesion phenomena we also demonstrated that when the gel is slightly compressed, the concentration of the surface tends to be very close to the concentration in bulk. This means that for a meaningful evaluation of the adhesive properties of hydrogels with a contact test, the gel should be slightly compressed during the formation of the contact. These results stress the need of a careful control of the polymerization conditions and the environment when synthesizing gels by simultaneous polymerization and crosslinking.

# ACKNOWLEDGEMENTS

We gratefully acknowledge the ANR Blanc Programme: project ADHGEL for its financial support.

## SUPPORTING INFORMATION AVAILABLE

A reflectivity curve displayed on the whole range of wave vectors with its best fit and the corresponding volume fraction profile of the PAA brush are presented in Supporting Information. This material is available free of charge via the Internet at http://pubs.acs.org.